\documentclass[%
a4paper,       
UKenglish,     
DIV10,          
useAMS,usenatbib,referee, doublespace]%
{scrartcl}

\usepackage[utf8]{inputenc}
\usepackage[UKenglish]{babel}
\usepackage[T1]{fontenc}
\usepackage{textcomp}
\usepackage{setspace}
\usepackage{hyperref}  
\usepackage{url}                
\usepackage[sumlimits, intlimits, namelimits]{amsmath} 
\usepackage{amssymb}            
\usepackage[caption = true]{subfig} 
\usepackage{array}              
\usepackage{booktabs}           
\usepackage{ae}                 
\usepackage[longnamesfirst,round]{natbib} 
\usepackage{color}              
\usepackage[sc]{mathpazo}       
\usepackage{helvet}             
\usepackage{todonotes}          

\hypersetup{
  bookmarksopen=true, 
  breaklinks=true,
  pdftitle={$p$-values for credibility},
  pdfsubject={Statistics},
  pdfkeywords={Statistics, Evidence},
  colorlinks=true,
  linkcolor=blue,
  anchorcolor=black,
  citecolor=teal,
  urlcolor=purple
}

\usepackage{xifthen}            
\usepackage{array}
\usepackage{amssymb}

\newcommand{\blanco}[1]{  } 

\newcommand{\latin}[1]{\textit{#1}}

\newcommand{\abk}[1]{\mbox{#1}\xdot}
\DeclareRobustCommand\xdot{\futurelet\token\Xdot}
\def\Xdot{%
  \ifx\token\bgroup.%
  \else\ifx\token\egroup.%
  \else\ifx\token\/.%
  \else\ifx\token\ .%
  \else\ifx\token!.%
  \else\ifx\token,.%
  \else\ifx\token:.%
  \else\ifx\token;.%
  \else\ifx\token?.%
  \else\ifx\token/.%
  \else\ifx\token'.%
  \else\ifx\token).%
  \else\ifx\token-.%
  \else\ifx\token+.%
  \else\ifx\token~.%
  \else\ifx\token.%
  \else.\ %
  \fi\fi\fi\fi\fi\fi\fi\fi\fi\fi\fi\fi\fi\fi\fi\fi%
}

\newcommand{\ie}{\abk{\latin{i.\,e}}}

\newlength{\halbebreite}
\DeclareMathOperator{\Nor}{N} 
  
\DeclareMathOperator{\Be}{Be} 

\newcommand{\partialv}[3][1]{%
\ifthenelse{#1 = 1}{\frac{\partial #2}{\partial #3}}{\frac{\partial^{#1} #2}{\partial #3^{#1}}}
} 

\newcommand{\partials}[3][1]{%
\ifthenelse{#1 = 1}{\frac{d #2}{d #3}}{\frac{d^{#1} #2}{d #3^{#1}}}
} 

\newcommand{\dseps}[2][1]{%
\ifthenelse{#1 = 1}{\frac{d}{d #2}}{\frac{d^{#1}}{d #2^{#1}}}
}

\newcommand{\dsepv}[2][1]{%
\ifthenelse{#1 = 1}{\frac{\partial}{\partial #2}}{\frac{\partial^{#1}}{\partial #2^{#1}}}
}

\newcommand{\ml}[2][1]{
\ifthenelse{#1 = 1}%
 {\hat{#2}_{\scriptscriptstyle{\mathrm{ML}}}}%
 {\hat{#2}^{#1}_{\scriptscriptstyle{\mathrm{ML}}}}
}
\newcommand{\map}[2][0]{
\ifthenelse{#1 = 0}%
 {\hat{#2}_{\scriptscriptstyle{\mathrm{MAP}}}}%
 {\hat{#2}_{{\scriptscriptstyle{\mathrm{MAP}}_{#1}}}}
}
\newcommand{\mpm}[2][0]{
\ifthenelse{#1 = 0}%
 {\hat{#2}_{\scriptscriptstyle{\mathrm{MPM}}}}%
 {\hat{#2}_{{\scriptscriptstyle{\mathrm{MPM}}_{#1}}}}
}

\newcommand{\given}{\,\vert\,} 

\newcommand{\abs}[1]{\left\lvert#1\right\rvert} 

\IfFileExists{upquote.sty}{\usepackage{upquote}}{}

\usepackage{Sweave}
\begin{document}

\title{$p$-Values for Credibility} 

\author{Leonhard
  Held\\ 
  Epidemiology, Biostatistics
  and Prevention Institute (EBPI)\\ University of Zurich\\ Hirschengraben 84,
  8001 Zurich, Switzerland\\ Email: \texttt{leonhard.held@uzh.ch}}

\maketitle

\begin{center}
\begin{minipage}{12cm}
\textbf{Abstract}: Analysis of credibility is a reverse-Bayes technique
that has been proposed by Matthews (2001) to overcome some of the
shortcomings of significance tests.  A significant result is deemed
credible if current knowledge about the effect size is in conflict
with any sceptical prior that would make the effect non-significant.
In this paper I formalize the approach and propose to
use Bayesian predictive tail probabilities to quantify the evidence
for credibility. This gives rise to a $p$-value for extrinsic
credibility, taking into account both the internal and the external
evidence for an effect. The assessment of intrinsic
credibility leads to a new threshold for ordinary significance that is
remarkably close to the recently proposed 0.005 level. Finally, a
$p$-value for intrinsic credibility is proposed that is a simple
function of the ordinary $p$-value for significance and has a direct
frequentist interpretation in terms of the replication probability
that a future study under identical conditions will give an estimated
effect in the same direction as the first study. 
\end{minipage}
\end{center}

\section{Introduction}
\begin{flushright}
``$p$-values are just too familiar and useful to ditch''\\
  David Spiegelhalter (2017)

\end{flushright}
Standard $P$-values for
point null hypotheses still dominate most of the applied literature
\citep{GreenlandPoole:2013}, despite the fact that they are commonly
misused and misunderstood
\citep{WassersteinLazar:2016,matthews2017}. Although being criticised 
intensively in the literature, the 
dichotomisation of $P$-values 
into ``significant'' and ``non-significant'' is still commonplace in practice. 

In a series of papers, Robert Matthews
\cite{matthews:2001,matthews:2001b,matthews:2017} has developed the Analysis of
Credibility, a specific
reverse-Bayes method to assess the credibility of a significant
finding. 
Reverse-Bayes approaches allow to study properties of the prior distribution needed to
achieve a certain posterior statement for the data at hand. 
The idea to use Bayes's theorem in reverse originates in the work
by IJ Good \citep{good:1983} and is increasingly used to assess
the plausibility of scientific findings
\citep{greenland:2006,greenland:2011,held2013,Colquhoun:2017}. 

Analysis of credibility is based on a conventional confidence interval
for an unknown effect size $\theta$ with lower limit $L$ and upper
limit $U$, say.  In the following I assume that both $L$ and $U$ are
symmetric around the point estimate $\hat \theta$ (assumed to be
normally distributed) and that both are either positive or negative,
\ie the effect is significant at significance level $\alpha$.
\citet{matthews:2001,matthews:2001b} proposed to compute a sceptical
prior distribution for the effect size $\theta$, normal with mean
zero, that - combined with the information given in the confidence
interval for $\theta$ - results in a posterior distribution which is
just non-significant at level $\alpha$, \ie either the $\alpha/2$ or
the $1-\alpha/2$ posterior quantile is zero.  He has derived a formula
for the limits $\pm S$ of the corresponding equi-tailed prior credible
interval at the same level $1-\alpha$:
\begin{equation}\label{eq:S}
S = \frac{(U-L)^2}{4 \sqrt{UL}},
\end{equation}
where $S$ is called the {\em sceptical limit} and the interval
$[-S,S]$ is called the {\em critical prior interval}.  Note that
\eqref{eq:S} holds for any value of $\alpha$, not just for
the traditional 5\% level.

Equivalently, the variance $\tau^2$ of the sceptical prior can be
expressed as a function of the variance $\sigma^2$ (the squared
standard error, assumed to be known) of the estimate $\hat \theta$,
the corresponding test statistic $t=\hat \theta/\sigma$ and
$z_{\alpha/2}$, the $1-\alpha/2$ quantile of the standard normal
distribution:
\begin{equation}\label{eq:tau2}
\tau^2 = \frac{\sigma^2}{t^2/z_{\alpha/2}^2 - 1},
\end{equation}
where $t^2>z_{\alpha/2}^2$ is required for significance at level $\alpha$.
Equation \eqref{eq:tau2} shows that the prior variance $\tau^2$ can be
both smaller or larger than $\sigma^2$, depending on the value of $t^2$. If
$t^2$ is close to $z_{\alpha/2}^2$ (\ie the effect is ``borderline
significant''), then the prior variance will be relatively large. If $t^2$
is substantially larger than $z_{\alpha/2}^2$, then the prior variance will be
relatively small.

\begin{center}
\begin{figure}[ht]
\begin{center}
\includegraphics{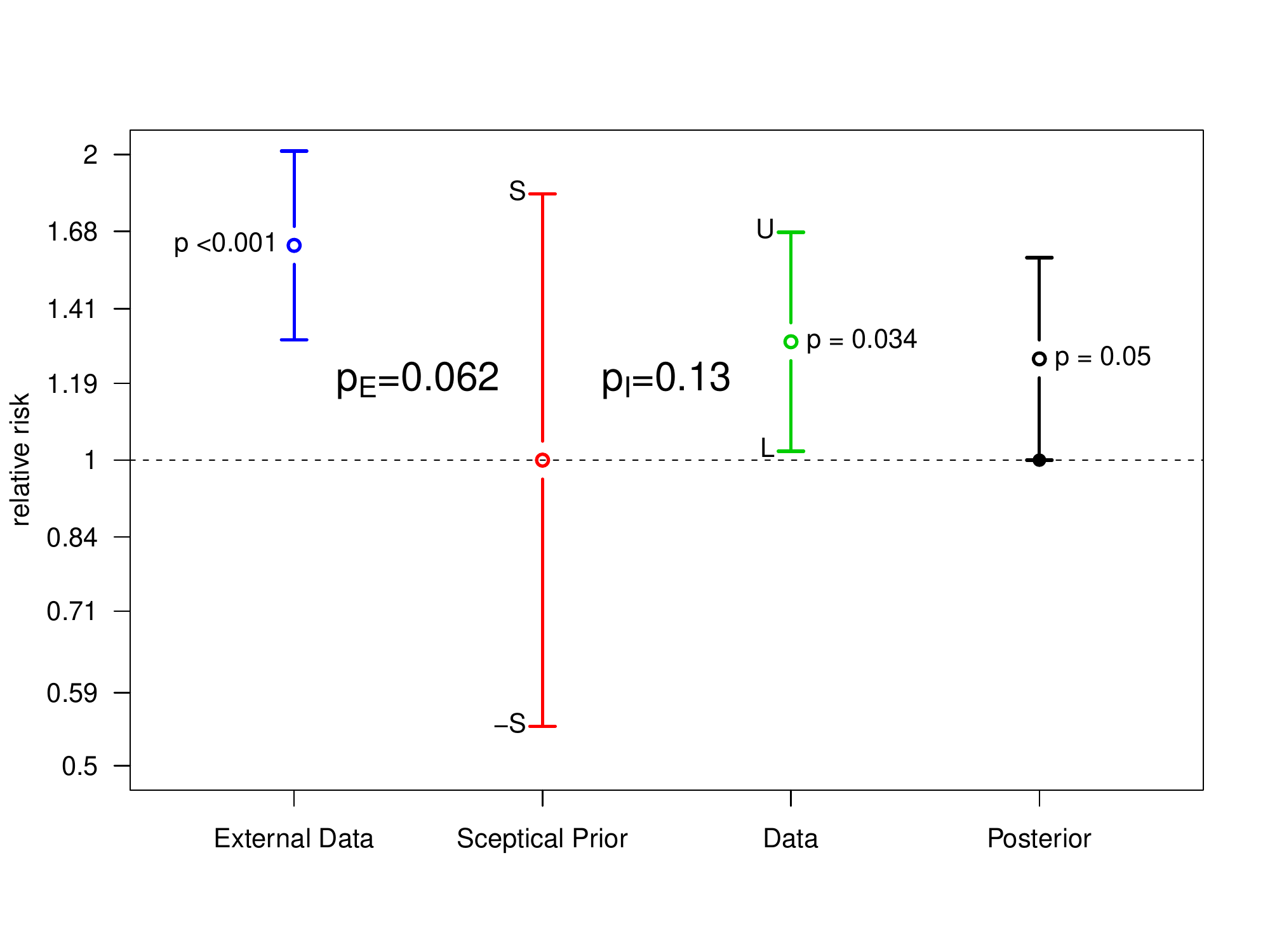}
\caption{Analysis of credibility for the relative risk of complete resolution of pain at
48 hours as reported in the \citet{Hayward_etal:2017} trial. The sceptical prior
for the relative risk $\theta$ has 95\% credible interval from 0.55 to 1.83. Combined with the data represented by the 95\% confidence interval from 1.02 to 1.68, the resulting equi-tailed 95\% posterior credible interval has lower limit 1. The external data is based on a meta-analysis of preceding trials. 
The $p$-value $p_E$ for extrinsic credibility is 
discussed in Section \ref{sec:sec2.2}. The $p$-value $p_I$ for intrinsic credibility is 
discussed in Section \ref{sec:sec3.2}. 
\label{fig:fig1}}
\end{center}
\end{figure}
\end{center}

For illustration, consider results from a recently published randomized
placebo-controlled clinical trial \citep{Hayward_etal:2017} on the
efficacy of corticosteroids in the treatment of sore throat. There
were 102/288 events in the intervention
group and 75/277 events in the control
group for the outcome complete resolution of pain at 48 hours. A
relative risk of $\exp(\hat \theta) = 1.31$ can easily be calculated (95\% CI from 1.02 to 1.68, $p=0.034$).
Analysis of credibility for the log relative risk $\theta$
gives the sceptical limit
0.60 = $\log(1.83)$, so
the critical prior interval is from $1/1.83 = 0.55$ to
1.83 on the relative risk scale.
The associated standard deviation $\tau=0.31$ of the sceptical prior is
considerably larger than the standard error 
$\sigma=0.13$.
Figure \ref{fig:fig1} displays the sceptical prior together with the 
confidence interval for the data and the associated posterior. 










\citet{matthews:2001b} proposed to compare external
knowledge about effect sizes with the sceptical limit $S$:
\begin{center}
\begin{minipage}{12cm}
{\em If previous evidence indicates that plausible values for the
  parameter in question exist outside the critical prior interval
  (CPI), the reality of the stated effect may be deemed credible [$\ldots$].}
\end{minipage}
\end{center}
In the above example, extrinsic credibility of the
\citet{Hayward_etal:2017} results can be investigated in the light of
preceding trials on the same clinical research
question. \cite{Sadeghiradj3887} performed a systematic review and
identified three preceding studies relevant for the
\citet{Hayward_etal:2017} analysis. 
A random-effects meta-analysis of the three trials preceding the Hayward (2017) study gives a
relative risk estimate of 
$1.63$ (95\% CI
from 1.31 to 2.02, $p < 0.0001$), also shown in Figure   \ref{fig:fig1}.
Since the external point estimate is smaller than the sceptical limit 1.83, the results
from the \citet{Hayward_etal:2017} study are not considered credible (at the 95\% confidence level) using the \citet{matthews:2001b} approach.


This approach seems somewhat unsatisfactory, since 
the dichotomisation into credible or not credible seems too
simplistic, just as the dichtomisation into significant and
non-significant. Instead, a quantitative
measure of credibility seems warranted. 
To this end I propose a more formal
assessment of credibility based on the \citet{box:1980} proposal to
quantify prior-data conflict with a Bayesian tail probability (Section \ref{sec:extrinsic}).  
This gives rise to a $p$-value for extrinsic
credibility, taking into account both the internal and the external
evidence for a significant effect, see Section \ref{sec:sec2.2}. 
The assessment of intrinsic credibility is described in Section
\ref{sec:sec3}, where a new justification for the recently proposed
0.005 threshold for significance is given (Section
\ref{sec:sec3.1}). A new $p$-value for intrinsic credibility is introduced
in Section \ref{sec:sec3.2} and interpreted as a replication
probability in Section \ref{sec:sec3.3}. I investigate the distribution of
$p$-values for credibility under the assumption of no effect (Section \ref{sec:sec4}), 
and close with some
comments in Section \ref{sec:sec6}.

\section{Assessing evidence for extrinsic credibility}
\label{sec:extrinsic}


In the present context the prior is the sceptical prior and the data
are derived from the external study used to assess credibility.  The
sceptical prior is normal with mean zero and variance $\tau^2$. The
likelihood of the external information on the effect size $\theta$ can
usually be described as normal with mean $\hat \theta_0$ and variance
$\sigma_0^2$, say. The \citet{box:1980} approach is now based on the
prior-predictive distribution, which is normal with mean zero and
variance $\tau^2 + \sigma_0^2$ \citep[Section 5.8]{sam:2004}.  The
procedure computes the test statistic
\begin{equation}\label{eq:tbox}
t_{\mbox{\scriptsize Box}} = \frac{\hat \theta_0}{\sqrt{\tau^2+\sigma_0^2}}
\end{equation}
and finally the (two-sided) tail probability 
\begin{equation}\label{eq:box.p}
  p_{\mbox{\scriptsize Box}} =\Pr(\chi^2(1) \geq t_{\mbox{\scriptsize Box}}^2)
  \end{equation}
as the corresponding upper tail of a
$\chi^2$-distribution with one degree of freedom. 
Small values of $p_{\mbox{\scriptsize Box}}$ indicate a conflict between the sceptical prior
and the external data. 
Note that this procedure is different from the assessment
of compatibility of the current study with the external study - then
we would use the test statistic $({\hat \theta_0- \hat \theta})/
{\sqrt{\sigma^2+\sigma_0^2}}$, which will give small tail
probabilities whenever the results from the two studies are
incompatible, independent of whether the trials show evidence for an
effect or not.

The proposed assessment of conflict between the sceptical prior
derived from the 95\% confidence interval and the meta-analytic
summary estimates from the preceding three trials gives
$t_{\mbox{\scriptsize Box}}= 1.49$ with Box's tail
probability $p_{\mbox{\scriptsize Box}}=0.14$.
However, this tail probability depends on 
$\alpha$ through $\tau^2$ as given in \eqref{eq:tau2}, here 5\%. 
For example, for $\alpha=10\%$, Box's tail probability is considerably smaller 
($p_{\mbox{\scriptsize Box}}=0.004$).  This is
because the lower limit of the 90\% confidence interval is further
away from the null than the lower limit of the 95\% confidence
interval, so we need a smaller prior variance $\tau^2$ to push it to
zero. A smaller $\tau^2$ results in a larger test statistic
\eqref{eq:tbox}, hence a smaller tail probability \eqref{eq:box.p}. In
practice, however, it is difficult to interpret tail probabilities for
credibility that depend on the confidence level $1-\alpha$ of the
underlying confidence interval.  Furthermore, computation of
$p_{\mbox{\scriptsize Box}}$ is not possible if the result from the
underlying study is not significant at level $\alpha$. These
issues motivate the work described in the next section where I define
a $p$-value for extrinsic credibility, independent of the level $\alpha$.

\subsection{A $p$-value for extrinsic credibility}
\label{sec:sec2.2}
There are two disadvantages of the procedure described in the previous
section: First, the tail probability $p_{\mbox{\scriptsize Box}}$ for
extrinsic credibility depends on the confidence level $1-\alpha$ of
the underlying confidence interval. Secondly, $p_{\mbox{\scriptsize
    Box}}$ only exists if the confidence interval does not include
zero, \ie is significant at level $\alpha$.  To address both problems,
I suggest to determine the largest confidence level $1-\alpha$, where
the stated effect is (just) extrinsically credible at level
$1-\alpha$, in analogy to the usual assessment of significance based
on confidence intervals. The required level $\alpha$ will be
called the {\em $p$-value for extrinsic credibility}, denoted by $p_E$. 

To determine $p_E$, let $c = \sigma^2/\sigma_0^2$ denote the
ratio of the variances of the internal and external
effect estimates. With
\eqref{eq:tau2} we then have
\[
\tau^2 + \sigma_0^2 = \sigma_0^2 \left(\frac{c}{t^2/z_{\alpha/2}^2 - 1} + 1 \right).
\]
Using \eqref{eq:tbox}, the requirement $t_{\mbox{\scriptsize Box}}^2 \geq z_{\alpha/2}^2$ for extrinsic credibility is 
then equivalent to 
\begin{equation}\label{eq:extrinsic.p}
\left(\frac{t_0^2}{z_{\alpha/2}^2}-1 \right) \left( \frac{t^2}{z_{\alpha/2}^2} - 1 \right) \geq c, 
\end{equation}
here $t_0=\hat \theta_0/\sigma_0$ is the test statistic of the external
study.  The required level $\alpha=p_E$ to obtain equality in
\eqref{eq:extrinsic.p} can easily be computed numerically. 
Note that \eqref{eq:tau2} requires $t^2>z_{p_E/2}^2$ to hold, so
also $t_0^2>z_{p_E/2}^2$ must hold 
to have the left side of \eqref{eq:extrinsic.p} being positive.
In
other words, $p_E$ will be larger than both ordinary (two-sided)
$p$-values for significance $p = 2 \left[1-\Phi(t)\right]$ and $p_0 =
2 \left[1-\Phi(t_0)\right]$, say, from the internal and the external
data, respectively, here $\Phi(.)$ denotes the cumulative standard
normal distribution function.

\begin{center}
\begin{figure}[ht]

\includegraphics{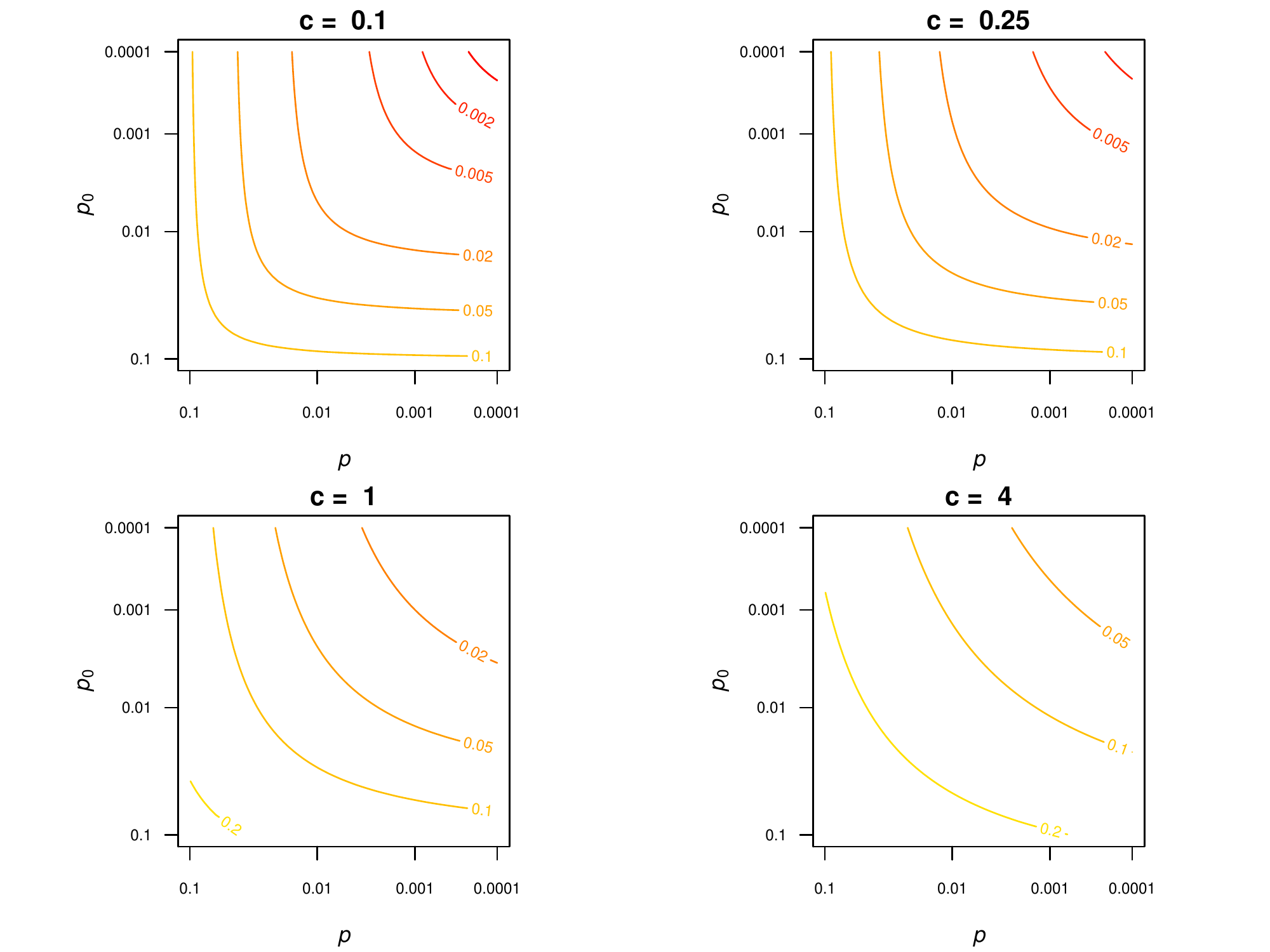}

\caption{The $p$-value for extrinsic credibility as a function of the ordinary $p$-values $p$ and $p_0$ of the internal and external studies. Shown are contour plots for different values of the variance ratio $c$.}
\label{fig:fig1b}
\end{figure}
\end{center}

Figure \ref{fig:fig1b} shows the dependence of $p_E$ on $p$, $p_0$ and
$c$.  Note that $p_E$ remains unchanged if we switch $t_0$ and $t$ but
keep $c$ fixed. This does {\em not} mean that switching the role of
the external and internal data will not change $p_E$ as then $c$ would
also change, except for the case $c=1$, where $\sigma^2=\sigma_0^2$ holds.
For fixed $t$ and $t_0$, $p_E$ is increasing with $c$, so the evidence
for extrinsic credibility decreases with increasing variance ratio
$\sigma^2/\sigma_0^2$ if the two test statistics $t$ and $t_0$ (and
the associated ordinary $p$-values) are kept constant. If $c$ is
small, then $p_E$ will be close to $\max\{p_0, p\}$, see Figure
\ref{fig:fig1b}. If $p_0 \ll p$, then $p_E$ will be close to $p$.

\setkeys{Gin}{width=1.0\textwidth}

In the above example we obtain $p_E=0.062$ by
numerical computation.
If we think of $p$-values as indicators of the strength of evidence 
and adopt the ``rough and ready
guide'' by \citet[Section 9.4]{Bland:2015}, then $p_E=0.062$ indicates weak evidence for credibility of the
results from the \citet{Hayward_etal:2017} trial in the light of the
three preceding trials.  A more technical interpretation is that for a
confidence level of $1-p_E=93.8$\% for the log
relative risk obtained from the \citet{Hayward_etal:2017} trial, Box's
tail probability \eqref{eq:box.p}, quantifying the conflict between
the corresponding sceptical prior and the external summary estimate, is equal to
$p_E=0.062$.

Note that the check ${\theta_0} > S$ for credibility by \cite{matthews:2001b} can be re-written as
$\theta_0^2/\tau^2 \geq z_{\alpha/2}^2$ and leads to 
the requirement
\begin{equation}\label{eq:extrinsic.p2}
\frac{t_0^2}{z_{\alpha/2}^2} \left(\frac{t^2}{z_{\alpha/2}^2} - 1 \right) \geq c.
\end{equation}
This is different from \eqref{eq:extrinsic.p} and no longer requires
${t_0^2}>{z_{\alpha/2}^2}$.  Specifically, if $c < 1$, it may happen that Matthews' check will declare a significant
result as extrinsically credible, although the external study was not
conventionally significant on its own right. 

\section{Assessing evidence for intrinsic credibility}
\label{sec:sec3}
A significant effect is intrinsically credible, if it is in conflict
with any sceptical prior that would make the effect non-significant.
This can be thought of as a double-check to ensure that
a significant effect is not spurious.
\citet{matthews:2017} proposed to declare an effect as
intrinsically credible, if the internal estimate $\hat \theta$ 
is outside the sceptical prior
interval $-S$ to $S$. He shows that,
for $\alpha=0.05$, this is equivalent to the conventional two-sided
$p$-value being smaller than 0.0127.

However, Matthews' check for intrinsic credibility does not take the
uncertainty of $\hat \theta$ directly into account. One could argue
that the sceptical prior variance \eqref{eq:tau2} is already a
function of the variance $\sigma^2$, so the uncertainty of $\hat
\theta$ is already implicitly taken into account in the calculation of
the sceptical limit.  This is true, but - for fixed $\sigma^2$ - the
point estimate $\hat \theta$ also affects the sceptical limit, so
Matthews' check does use $\hat \theta$ twice. It is not clear why
$\hat \theta$ can be used twice, but not $\sigma^2$.  In what follows
I will use both $\hat \theta$ and $\sigma^2$ directly in the
assessment of intrinsic credibility, following the approach by
\cite{box:1980} for the assessment of prior-data conflict, with the
perhaps unusual feature that the prior has been derived from the
data. I argue that there is nothing intrinsically inconsistent
in investigating the compatibility of a prior, defined through the data,
and the data itself, extending a argument made by \citet[Section
  5.10]{cox:2006} to the reverse-Bayes setting.


\subsection{Another justification for the 0.005 threshold}
\label{sec:sec3.1}

If we use the \citet{box:1980} check for prior-data
conflict between the sceptical prior and the original (internal) data with
point estimate $\hat \theta$ and variance $\sigma^2$, the test statistic
\eqref{eq:tbox} reads
\begin{equation}\label{eq:tbox2}
t_{\mbox{\scriptsize Box}}  = \frac{\hat \theta}{\sqrt{\tau^2+\sigma^2}}.
\end{equation}

Intrinsic credibility at the 5\% level (\ie $\abs{t_{\mbox{\scriptsize Box}}} > 1.96$) can
then be shown to be equivalent to the conventional two-sided
$p$-value $p<0.0056$.  This is remarkably close to the recently
proposed new threshold of 0.005 for statistical significance
\citep{Johnson2013,BenjaminEtAlinpress}.  

\begin{center}
\begin{figure}[ht]
\includegraphics{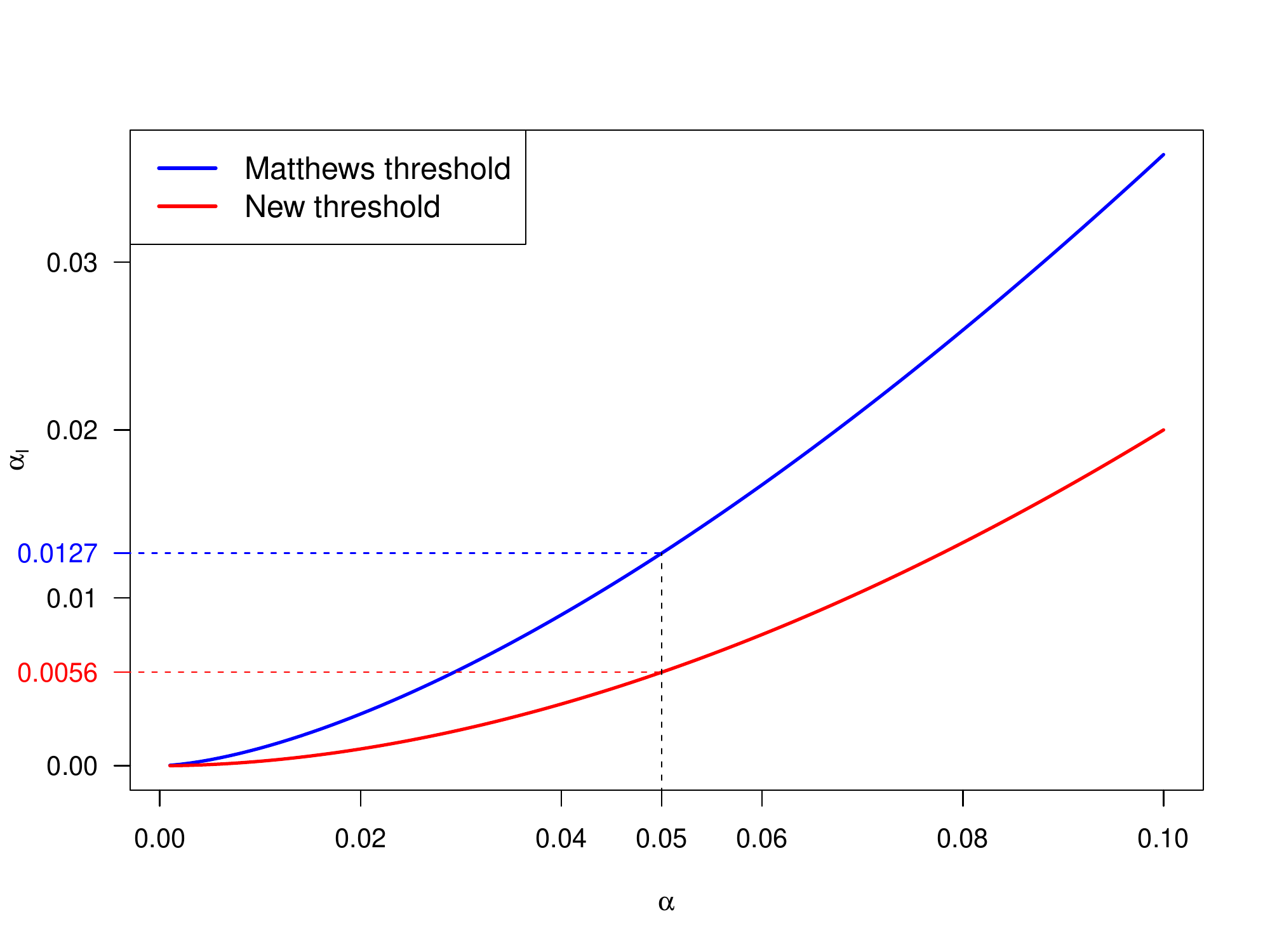}
\caption{The threshold for intrinsic credibility of significant results as a
  function of the conventional $\alpha$ level. The blue line
  corresponds to the proposal by \citet{matthews:2017}. The red line is the proposed new threshold.} 
\label{fig:fig2}
\end{figure}
\end{center}

To derive the new significance threshold, set $c=1$ and $t_0=t$ in \eqref{eq:extrinsic.p} 
to obtain the simple requirement 
\begin{eqnarray}\label{eq:int.cred}
  t^2 & \geq & 2 \, z_{\alpha/2}^2
\end{eqnarray}
for intrinsic credibility at level $\alpha$. 
In \eqref{eq:int.cred}, 
$t$ is the ordinary test statistic $t=\hat \theta/\sigma$, so 
the {\em intrinsic credibility threshold} $\alpha_I$ for the conventional two-sided $p$-value is
\begin{eqnarray}
\alpha_I &=& 2 \left\{1 - \Phi\left(t=\sqrt{2} \, z_{\alpha/2} \right) \right\}. \label{eq:iP}
\end{eqnarray}
For $\alpha=0.05$ we have $t=\sqrt{2}\cdot 1.96= 2.77$ and the
credibility threshold \eqref{eq:iP} turns out to be $\alpha_I =
0.0056$, as claimed above. 
Figure \ref{fig:fig2}
compares the new threshold with the one obtained by \citet[Appendix D]{matthews:2017}  (using 
$t=1.272 \, z_{\alpha/2}$) for values of $\alpha$ below 10\%.

The squared test statistic \eqref{eq:tbox2} can also be written in terms of $L$ and $U$, 
\[
t_{\mbox{\scriptsize Box}}^2 = z_{\alpha/2}^2 \frac{4 \, U L}{(U-L)^2},
\]
and the requirement $t_{\mbox{\scriptsize Box}}^2 \geq z_{\alpha/2}^2$ for 
intrinsic credibility can be shown to be equivalent to require
that the {\em credibility ratio} $U/L$ (or $L/U$ if both $L$ and $U$ are negative) fulfills
\begin{equation}\label{eq:cr}
\frac{U}{L} \leq d = 3 + 2 \, \sqrt{2} \approx 5.828.
\end{equation}
Thus, there is a simple way to assess intrinsic credibility based on
the ratio of the limits of a confidence interval at any level
$1-\alpha$, without using the ordinary $p$-value for significance: if
the credibility ratio is smaller than $d=5.828$ than the
result is credible at level $1-\alpha$. Note that this is a stronger
requirement than the check of confidence intervals for significance,
where it is only required that the credibility ratio is positive.  To
derive the cut-point $d$ in \eqref{eq:cr}, set $U=L \, d$ so the
requirement $t_{\mbox{\scriptsize Box}}^2 = z_{\alpha/2}^2$ is
equivalent to
$$1  =  \frac{4 \, U L}{(U-L)^2} =  \frac{4 \, d}{(d-1)^2},$$ 
a quadratic equation in $d$ with $d=3 + 2 \, \sqrt{2}$ as solution.

If the sceptical prior distribution is already available, then another
way to assess intrinsic credibility is to compare the prior variance
$\tau^2$ to the data variance $\sigma^2$. Comparing \eqref{eq:tau2}
with \eqref{eq:int.cred} it is easy to see that 
intrinsic credibility is achieved if and only if the sceptical prior variance is not larger than the variance of the estimate, \ie $\tau^2 \leq
\sigma^2$.

\subsection{A $p$-value for intrinsic credibility}
\label{sec:sec3.2}
As for the $p$-value for extrinsic credibility introduced in Section \ref{sec:sec2.2}, 
I now derive the $\alpha$ value that just achieves
intrinsic credibility, \ie where equality holds in \eqref{eq:int.cred}.
This gives the {\em $p$-value for intrinsic
  credibility}, denoted by $p_I$. The
$p$-value for credibility will always be larger than the $p$-value for
significance, which is based on the confidence level $1-\alpha$ such
that the lower limit $L$ is exactly zero.

The $p$-value $p_I$ for intrinsic
credibility can be written as a function of the ordinary
$p$-value $p$ for significance:
\begin{equation}\label{eq:eq3}
p_I=2\left[1-\Phi\left(t/\sqrt{2} \right)\right] ,
\end{equation}
here $t=t(p)= \Phi^{-1}(1-p/2)$ is the standard test statistic for significance. Equation \eqref{eq:eq3} can be derived by solving equation
\eqref{eq:iP} for $\alpha = p_I$ and replacing $\alpha_I$ with $p$.  Note that the corresponding test statistic $t_I = t / \sqrt{2}$ for
intrinsic credibility is just a shrunken version 
of the test statistic $t$ for significance.

\begin{center}
\begin{figure}[ht]
\includegraphics{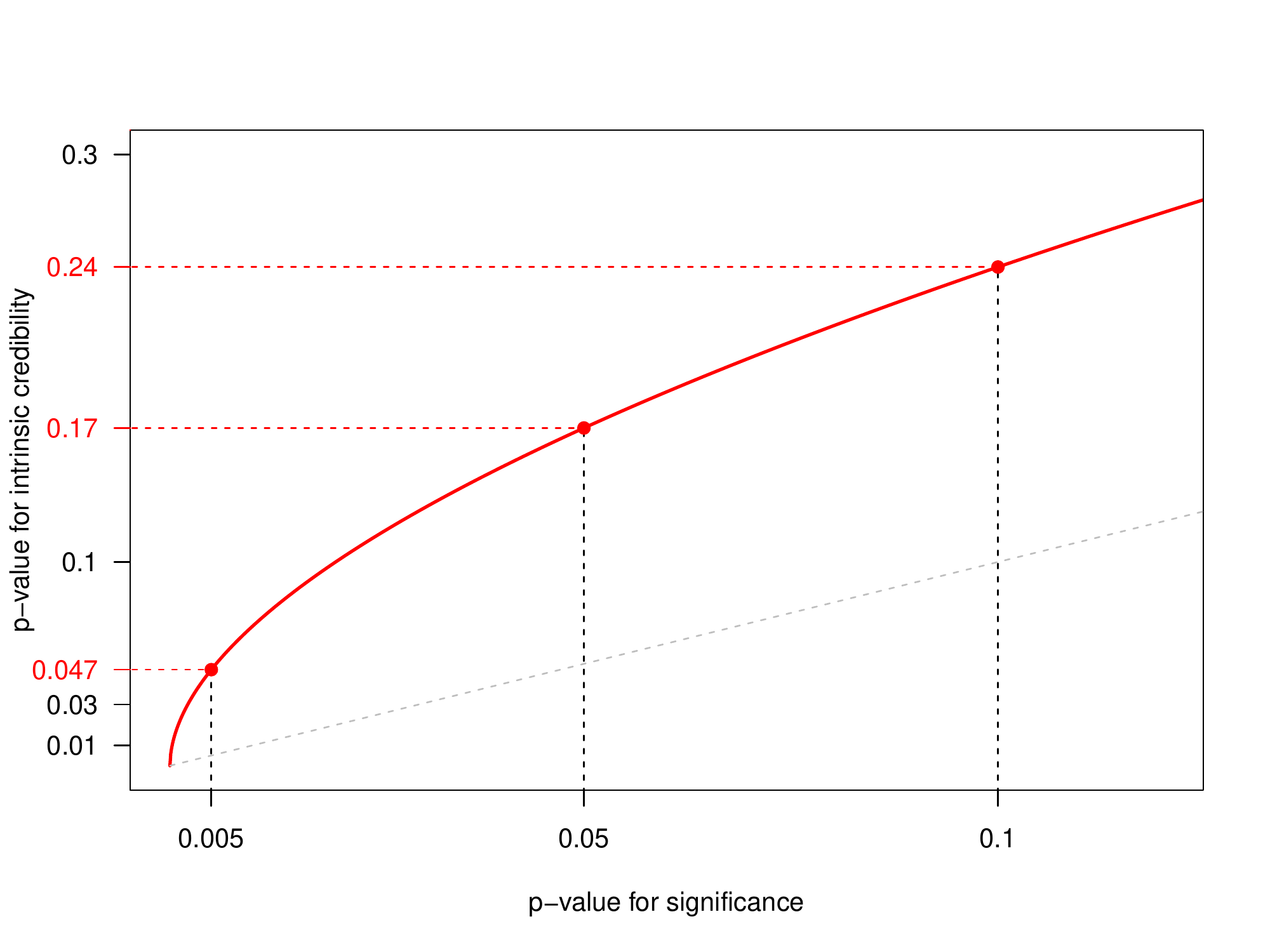}
\caption{The $p$-value for intrinsic
credibility as a function of the $p$-value for
significance. The grey dashed line is the identity line.}
\label{fig:fig3}
\end{figure}
\end{center}

Figure \ref{fig:fig3} shows that the $p$-value $p_I$ for intrinsic 
credibility is considerably larger than the $p$-value $p$ for significance, particularly
for small values of $p$. 
For example, the $p$-value for intrinsic
credibility of the \citet{Hayward_etal:2017} study is
$p_I = 0.13$, whereas the
conventional $p$-value for significance is $p=0.034$.


\subsection{Interpretation as replication probability}
\label{sec:sec3.3}
The $p$-value for intrinsic credibility has a direct and useful
interpretation in terms of the replication probability that an
identical study will give an estimated effect $\hat \theta_2$ in the
same direction as the estimate $\hat \theta_1 = \hat \theta$ from the current
(first) study. To see this, note that under an initial uniform prior
the posterior for $\theta$ is $\theta \given \hat \theta_1 \sim \Nor(\hat \theta_1,
\sigma^2)$. This posterior now serves as the prior for the mean of the (unobserved) estimate
$\hat \theta_2 \given \theta \sim \Nor(\theta, \sigma^2)$ from the second study,
where we assumed the two studies to be identically designed, so with equal
variances $\sigma^2$.  This leads to the prior-predictive distribution
$\hat \theta_2  \given \hat \theta_1 \sim \Nor(\hat \theta_1, 2 \, \sigma^2)$ and the $p$-value 
for intrinsic credibility \eqref{eq:eq3} is then
twice the probability that the second study will give an 
estimate $\hat \theta_2$ in the opposite direction as the estimate $\hat \theta_1$ of the 
first study:
\begin{eqnarray*}
2\, \Pr(\hat \theta_2 \leq 0 \given \hat \theta_1 > 0) &=& 2\, \Phi\left(\frac{0 - \hat \theta_1}{\sqrt{2}\sigma} \right) \\
&=& 2\, \Phi\left({-t}/{\sqrt{2}} \right) \\
&=& 2\, \left[1-\Phi\left({t}/{\sqrt{2}} \right) \right] \\
&=& p_I,
\end{eqnarray*}
and vice versa if $\hat \theta_1 < 0$.
The probability $\Pr(\hat \theta_2 \leq 0 \given \hat \theta_1 > 0)$
is one of the three replication
probabilities that have been considered by \citet{SIM:SIM1072}. A
related, but different quantity has been calculated by
\citet{SIM:SIM4780110705}, the probability that the result from the
second study is significant.

\section{The distribution under the null}
\label{sec:sec4}

\begin{center}
\begin{figure}[!ht]
\includegraphics{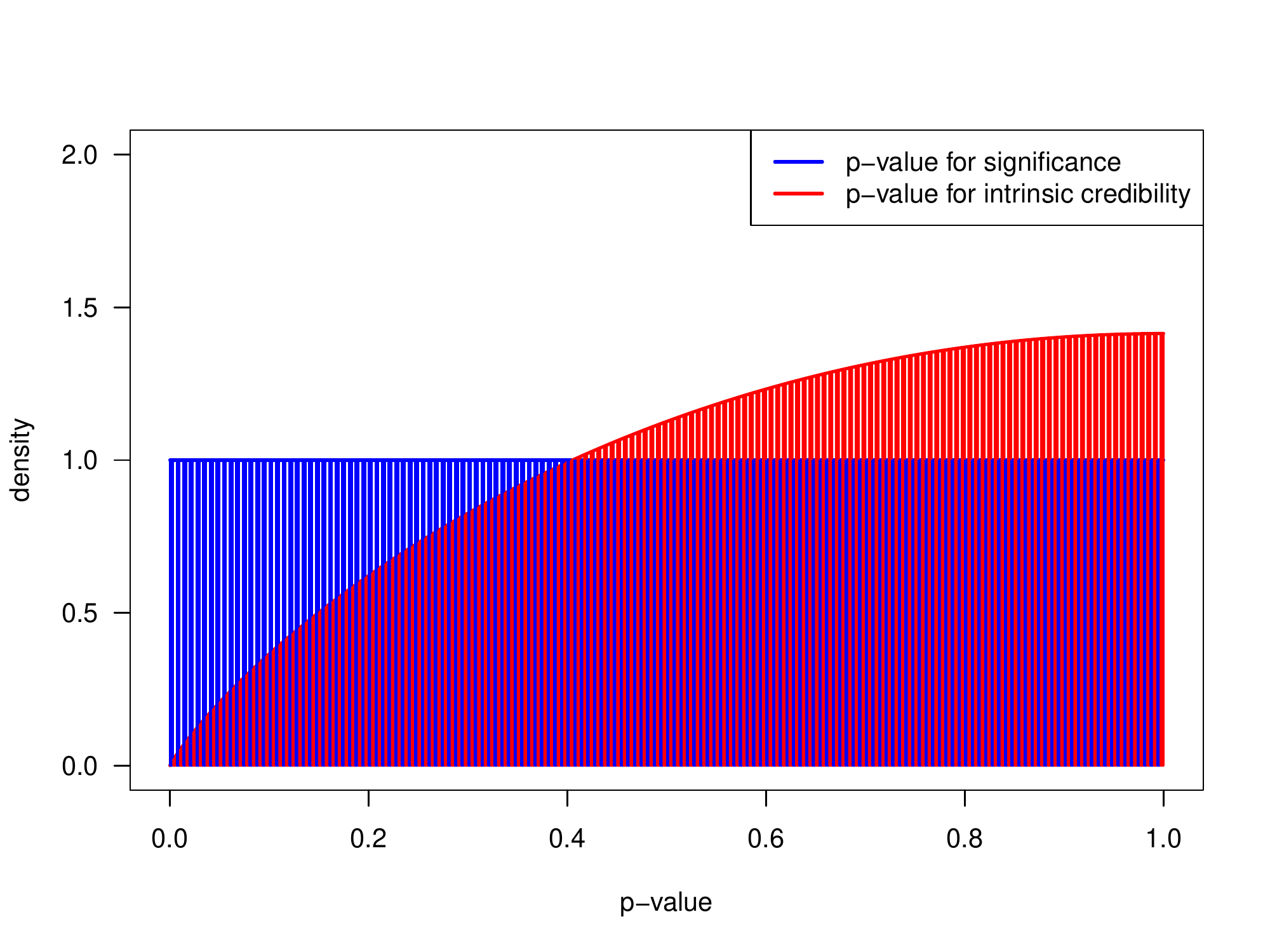}
\caption{The density function of the $p$-value for significance 
  and the $p$-value for intrinsic credibility under the assumption of no effect. 
}
\label{fig:fig4}
\end{figure}
\end{center}

It is of interest to study the distribution of $p_I$ and $p_E$ under the
assumption of no effect. Then $p$ is uniformly distributed and we can 
derive the density of $p_I$ with a
change-of-variables using \eqref{eq:eq3}:
\[
f(p_I) = 2 \sqrt{\pi} \, \varphi\left\{t(p_I)\right\},
\]
here $\varphi(.)$ is the standard normal density function and
$t(p_I)=\Phi^{-1}(1- p_I/2)$.  The density functions of $p$ and $p_I$ are
compared in Figure \ref{fig:fig4}. Under the assumption of no effect, the risk of small $p$-values
for intrinsic credibility is drastically reduced, compared to standard $p$-values for significance. 

\begin{figure}[!ht]
\begin{center}
\includegraphics{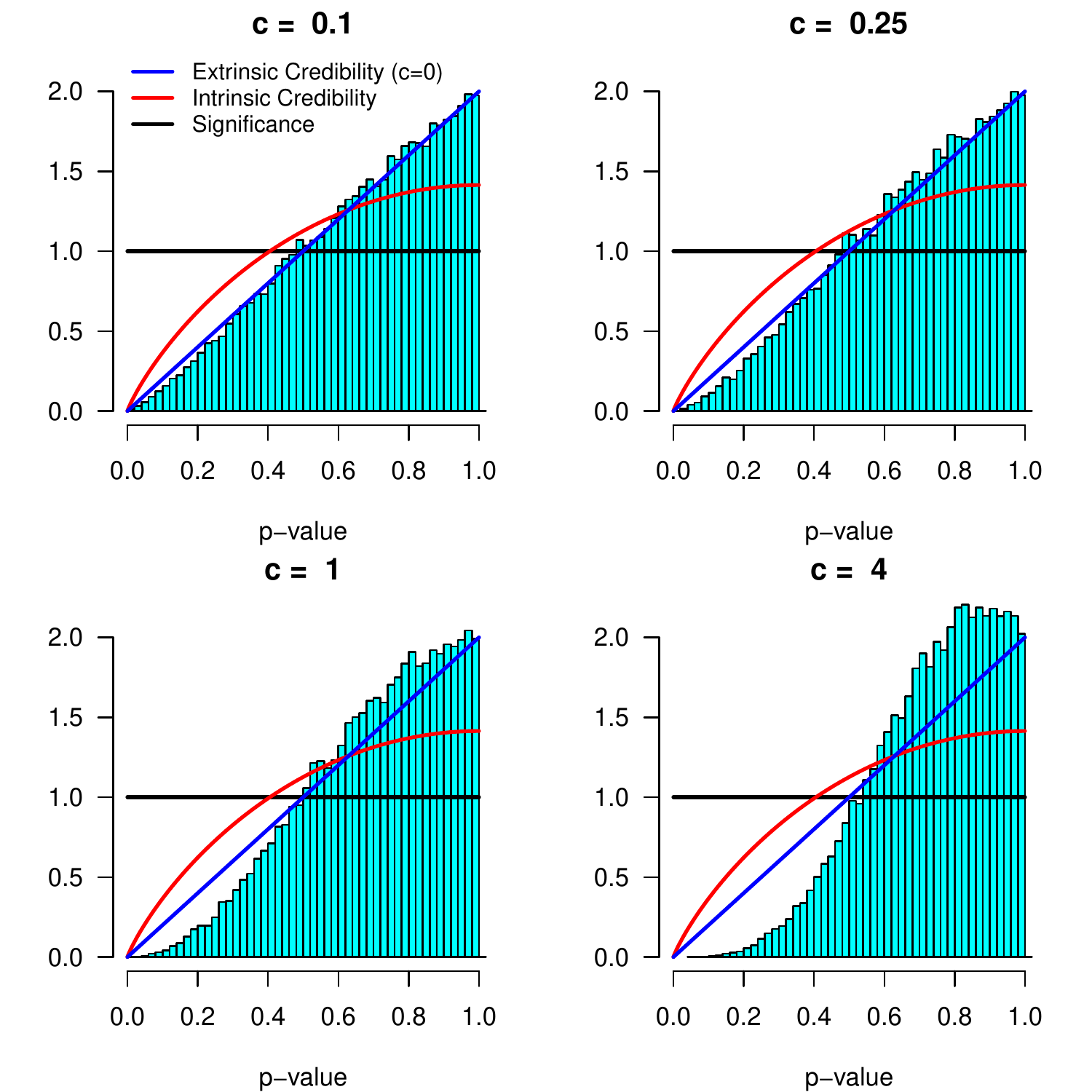}
\end{center}
\caption{Histograms of the distribution of the $p$-value for extrinsic
  credibility for different values of $c$ under the assumption of no
  effect.  The density function of the $p$-value for extrinsic
  credibility for the limiting case $c=0$ is superimposed (blue line), as well as
  the density function of the $p$-value for intrinsic credibility (red line) and
  for significance (black line).}
\label{fig:fig5}
\end{figure}

The distribution of $p_E$ can be easily studies via simulation. 
Histograms of 50\,000 samples is displayed in Figure \ref{fig:fig5}
for the same values of $d$ as in Figure \ref{fig:fig5}. 
The
distribution of $p_E$ is shifted to the right with increasing $c$. For
$c \rightarrow 0$ we have $p_E \rightarrow \min\{p, p_0\}$, which
follows a rectangular $\Be(2, 1)$ distribution under the assumption
that $p$ and $p_0$ are independently uniform.   For
comparison, Figure \ref{fig:fig5} also gives the density function of
the $p$-value for significance and for intrinsic credibility. Under
the assumption of no effect, the risk of small ``false positive'' $p$-values for
credibility is drastically reduced, compared to standard $p$-values
for significance. This is already to see for $p_I$, but even more pronounced for
$p_E$. The rectangular
distribution (for $c \rightarrow 0$) gives an upper bound for the tail probability $\Pr(p_E <
\alpha \given H_0) \leq \alpha^2$ for any threshold $\alpha<0.5$ and
for any value of the variance ratio $c$. For example, for
$\alpha=0.05$ we obtain $\Pr(p_E < 0.05 \given H_0) \leq 0.0025$. For comparison, 
$\Pr(p_I < 0.05 \given H_0) = 0.0056$.

\section{Discussion}
\label{sec:sec6}

In this paper I have introduced $p$-values to assess extrinsic and
intrinsic credibility. The $p$-value for extrinsic credibility is a
function of the two ordinary $p$-values from the internal and
external data and the ratio of the squared standard errors of the
internal and external effect estimates only.  The $p$-value for intrinsic
credibility has been shown to be function of the ordinary $p$-value
for significance from the internal data.

The proposed $p$-values for credibility are always larger than the
corresponding $p$-values for significance. As such, they provide a new
calibration of ordinary $P$-values.  For the conventional
$\alpha=0.05$ level, the threshold for intrinsic credibility turns out
to be remarkably close to the recently proposed 0.005 significance level, which has
been proposed based on different arguments.

Although derived using a Bayesian approach, $p$-values for credibility
do not require specification of a prior probability of the null
hypothesis of no effect. In practice it is therefore easy to accompany
$p$-values for significance with the corresponding $p$-values for
credibility.  This is in contrast to the calibration of $p$-values to
lower bounds on posterior probability of the null, which require
specification of a prior probability.  However, it is noteworthy that
the $p$-value for intrinsic credibility is surprisingly close to a
commonly used bound on the posterior probability of the null
hypothesis (of no effect) \citep{sbb:2001} under the condition of
equipoise \citep{Johnson2013}.  Minimum Bayes factors have also been
proposed to calibrate $p$-values, see \citet{HeldOtt2018} for a recent
review. They have the advantage that they do not require
specification of a prior probability of the null hypothesis, but they
cannot be interpreted as probabilities. 

The analysis of credibility assumes a simple mathematical framework,
where likelihood, prior and posterior are all normally distributed. It
will be of interest to extend this approach to other settings, for
example to the $t$-distribution.

\section*{Acknowledgments}
I am indebted to Robert Matthews for inventing the Analysis of
Credibility, which forms the basis of the work presented here.  I am also
grateful to Stefanie Muff and Manuela Ott for proofreading this
manuscript.

\nocite{matthews2017}

\nocite{spiegelhalter2017}
\bibliographystyle{apalike}
\bibliography{antritt}

\begin{thebibliography}{}

\bibitem[Benjamin et~al., 2017]{BenjaminEtAlinpress}
Benjamin, D.~J., Berger, J.~O., Johannesson, M., Nosek, B.~A., Wagenmakers,
  E.-J., et~al. (2017).
\newblock Redefine statistical significance.
\newblock {\em Nature Human Behaviour}.
\newblock \url{http://dx.doi.org/10.1038/s41562-017-0189-z}.

\bibitem[Bland, 2015]{Bland:2015}
Bland, M. (2015).
\newblock {\em {An Introduction to Medical Statistics}}.
\newblock Oxford University Press, Oxford, 4th edition.

\bibitem[Box, 1980]{box:1980}
Box, G. E.~P. (1980).
\newblock Sampling and {B}ayes' inference in scientific modelling and
  robustness (with discussion).
\newblock {\em Journal of the Royal Statistical Society, Series A},
  {143}:383--430.

\bibitem[Colquhoun, 2017]{Colquhoun:2017}
Colquhoun, D. (2017).
\newblock The reproducibility of research and the misinterpretation of
  p-values.
\newblock {\em Royal Society Open Science}, 4(12).
\newblock \url{http://dx.doi.org/10.1098/rsos.171085}.

\bibitem[Cox, 2006]{cox:2006}
Cox, D.~R. (2006).
\newblock {\em Principles of Statistical Inference}.
\newblock Cambridge University Press, Cambridge.

\bibitem[Good, 1983]{good:1983}
Good, I.~J. (1983).
\newblock {\em {Good Thinking: The Foundations of Probability and Its
  Applciations}}.
\newblock University of Minnesota Press, Minneapolis.

\bibitem[Goodman, 1992]{SIM:SIM4780110705}
Goodman, S.~N. (1992).
\newblock A comment on replication, p-values and evidence.
\newblock {\em Statistics in Medicine}, 11(7):875--879.

\bibitem[Greenland, 2006]{greenland:2006}
Greenland, S. (2006).
\newblock Bayesian perspectives for epidemiological research: I. foundations
  and basic methods.
\newblock {\em International Journal of Epidemiology}, {35}:765--775.

\bibitem[Greenland, 2011]{greenland:2011}
Greenland, S. (2011).
\newblock Null misinterpretation in statistical testing and its impact on
  health risk assessment.
\newblock {\em Preventive Medicine}, {53}:225--228.

\bibitem[Greenland and Poole, 2013]{GreenlandPoole:2013}
Greenland, S. and Poole, C. (2013).
\newblock Living with p values: Resurrecting a {B}ayesian perspective on
  frequentist statistics.
\newblock {\em Epidemiology}, 24:62--68.

\bibitem[Hayward et~al., 2017]{Hayward_etal:2017}
Hayward, G., Hay, A., Moore, M., et~al. (2017).
\newblock Effect of oral dexamethasone without immediate antibiotics vs placebo
  on acute sore throat in adults: A randomized clinical trial.
\newblock {\em JAMA}, 317(15):1535--1543.

\bibitem[Held, 2013]{held2013}
Held, L. (2013).
\newblock {R}everse-{B}ayes analysis of two common misinterpretations of
  significance tests.
\newblock {\em Clinical Trials}, 10:236--242.

\bibitem[Held and Ott, 2018]{HeldOtt2018}
Held, L. and Ott, M. (2018).
\newblock On $p$-values and {B}ayes factors.
\newblock {\em Annual Review of Statistics and Its Application}, 5(1).

\bibitem[Johnson, 2013]{Johnson2013}
Johnson, V.~E. (2013).
\newblock Revised standards for statistical evidence.
\newblock {\em Proc. Natl. Acad. Sci. U.S.A.}, 110(48):19313--19317.

\bibitem[Matthews, 2001a]{matthews:2001}
Matthews, R. (2001a).
\newblock Methods for assessing the credibility of clinical trial outcomes.
\newblock {\em Drug Information Journal}, {35}:1469--1478.

\bibitem[Matthews, 2001b]{matthews:2001b}
Matthews, R. (2001b).
\newblock Why {\em should} clinicians care about {B}ayesian methods? (with
  discussion).
\newblock {\em Journal of Statistical Planning and Inference}, {94}:43--71.

\bibitem[Matthews, 2017]{matthews:2017}
Matthews, R. (2017).
\newblock {Beyond "significance": principles and practice of the Analysis of
  Credibility}.
\newblock {\em Royal Society Open Science}.
\newblock to appear.

\bibitem[Matthews et~al., 2017]{matthews2017}
Matthews, R., Wasserstein, R., and Spiegelhalter, D. (2017).
\newblock The {ASA}'s $p$-value statement, one year on.
\newblock {\em Significance}, 14(2):38--41.

\bibitem[Sadeghirad et~al., 2017]{Sadeghiradj3887}
Sadeghirad, B., Siemieniuk, R. A.~C., Brignardello-Petersen, R., et~al. (2017).
\newblock Corticosteroids for treatment of sore throat: systematic review and
  meta-analysis of randomised trials.
\newblock {\em BMJ}, 358.

\bibitem[Sellke et~al., 2001]{sbb:2001}
Sellke, T., Bayarri, M.~J., and Berger, J.~O. (2001).
\newblock Calibration of $p$ values for testing precise null hypotheses.
\newblock {\em The American Statistician}, 55:62--71.

\bibitem[Senn, 2002]{SIM:SIM1072}
Senn, S. (2002).
\newblock {Letter to the Editor: A comment on replication, p-values and
  evidence by S.N.~Goodman, Statistics in Medicine 1992; 11:875-879}.
\newblock {\em Statistics in Medicine}, 21(16):2437--2444.

\bibitem[Spiegelhalter, 2017]{spiegelhalter2017}
Spiegelhalter, D. (2017).
\newblock Too familiar to ditch.
\newblock {\em Significance}, 14(2):41.

\bibitem[Spiegelhalter et~al., 2004]{sam:2004}
Spiegelhalter, D.~J., Abrams, K.~R., and Myles, J.~P. (2004).
\newblock {\em {Bayesian Approaches to Clinical Trials and Health-Care
  Evaluation}}.
\newblock Wiley, New York.

\bibitem[Wasserstein and Lazar, 2016]{WassersteinLazar:2016}
Wasserstein, R.~L. and Lazar, N.~A. (2016).
\newblock The {ASA}'s statement on p-values: context, process, and purpose.
\newblock {\em Am. Stat.}, 70(2):129--133.

\end{thebibliography}
\end{document}